# Observation of magnetically coupled electro-optic effect in LiNbO$_3$/LiTaO$_3$ at room temperature

**Authors:** Yalong Yu[1], Hengwei Zhang[1], Xiaoyan Liu[1], and Tao Chu[1, *]

**Affiliations:**
[1] College of Information Science and Electronic Engineering, Zhejiang University, Hangzhou 310058, P. R. China

[*]Corresponding author. Email: chutao@zju.edu.cn;

**Abstract:** The magnetoelectric coupling effect serves as a crucial bridge between electrical and magnetic order parameters in condensed matter physics, forming the physical basis for the development of next-generation low-power information storage and sensing technologies. However, material systems exhibiting this effect at room temperature are extremely rare, and the coupling strength is typically very weak, which has long hindered the practical application of such phenomena despite their rich tunability. Here, we break this long-standing trade-off by reporting a pronounced magnetoelectric phenomenon—the magnetically coupled electro-optic effect—in the classic ferroelectric optical materials LiNbO$_3$ and LiTaO$_3$. We trace its origin to an unexpected source: the problematic "direct-current drift," a major reliability issue in photonic integrated circuits. We unambiguously demonstrate that this drift stems not from mobile ions but from defect-bound unpaired electrons, whose slow polarization relaxation is quenched upon the magnetic-field-induced formation of a room-temperature skyrmion states, as directly visualized by Lorentz transmission electron microscopy. This collective spin ordering not only solves the decades-old drift problem but also transforms the defect states into a magnetically responsive platform, exhibiting an efficiency up to 34 pm/V in LiNbO$_3$ and 15 pm/V in LiTaO$_3$—an orders-of-magnitude enhancement over conventional room-temperature magnetoelectric responses and even surpassing the materials' intrinsic Pockels coefficients. We also measured the magnetic response of this additional electro-optical effect and found that under the condition of applying a magnetic field of 0.1 T, an electro-optical coefficient adjustment of about 8 pm/V can be achieved (~0.008 pm/V/Oe), which is equivalent to ~ 30% of the electro-optical coefficient of LiNbO$_3$ itself. Our work bridges a hard-to-use device degradation phenomenon with the creation of a strong, room-temperature-stable magnetoelectrical coupling, opening avenues for non-volatile photonics and fundamentally new cross-correlation control in materials.

# Introduction

The search for multifunctional materials that exhibit cross-coupling between distinct ferroic orders, such as ferroelectricity and (anti)ferromagnetism, represents one of the cornerstones of modern condensed matter physics and materials science *(1-3)*. Among these, the magnetoelectric (ME) effect—where electric polarization can be induced by a magnetic field or magnetization by an electric field—holds significant promise for next-generation technologies, including low-power spintronic memories, magnetic field sensors, and electrically tunable microwave devices. Concurrently, the electro-optic (EO) effect, which enables the modulation of a material's optical properties via an electric field, forms the foundation of modern photonics and optical communications *(4-7)*. A compelling yet largely underexplored frontier lies at the intersection of these two phenomena: the possibility of magnetically tunable electro-optic responses. Such a capability would open the door to non-volatile, magnetically programmable photonic devices, effectively bridging the domains of spin control and light manipulation.

Despite its transformative potential, progress in this area has been substantially hindered by two fundamental and interrelated challenges. First, robust magnetoelectric coupling at room temperature is exceptionally rare *(8-10)*. Second, and even more critically, higher-order magneto-electro-optic coupling—where a magnetic field directly modulates the linear electro-optic (Pockels) coefficient—is a higher order effect described by a third-rank tensor *(11)*. This effect is inherently weaker and subject to more stringent symmetry restrictions than linear magnetoelectricity. As a result, it has remained largely a theoretical curiosity; unambiguous experimental observation, let alone achieving significant effect strength under room-temperature conditions, has been considered highly challenging, if not improbable. The prevailing view has been that any such coupling would be vanishingly small and observable only under extreme conditions of low temperature and high magnetic fields.

In this work, we experimentally demonstrate that a specific electrical phenomenon known as DC drift—which leads to performance degradation in devices based on ferroelectric materials with excellent optical properties and strong electro-optic effects, such as $LiNbO_3$ (LN) and $LiTaO_3$ (LT)—actually originates from a latent lattice-internal electronic response that can be modulated by a magnetic field, rather than from simple electrical relaxation of free charged carriers as previously assumed *(12-16)*. We further clarify the physical mechanism underlying direct-current (DC) drift and predict that it may give rise to an additional magnetically coupled electro-optic (mEO) effect. By suppressing DC drift through magnetic-field annealing, we measured an additional electro-optic contribution in Mach–Zehnder interferometer (MZI) modulators fabricated on lithium niobate-on-insulator (LNOI) and lithium tantalate-on-insulator (LTOI) platforms, with magnitudes reaching up to 80 pm/V in LN and 25 pm/V in LT. Through characterization of the magnetic-field response of this newly observed electro-optic effect, we confirm its significant magnetic tunability—that is, it constitutes a magnetically coupled electro-optic effect. These results break away from the prevailing perception that room-temperature magnetoelectric coupling is inherently extremely weak and reveal a novel pathway for exploring magnetoelectric phenomena. By uncovering a pronounced effect where it was least expected, our work establishes a new material platform for room-temperature magneto-optic device engineering. It provides direct experimental access to a fundamentally new class of higher-order cross-correlations and opens a window into the intricate interplay among spin, charge, and orbital degrees of freedom.

# Origin of DC drift

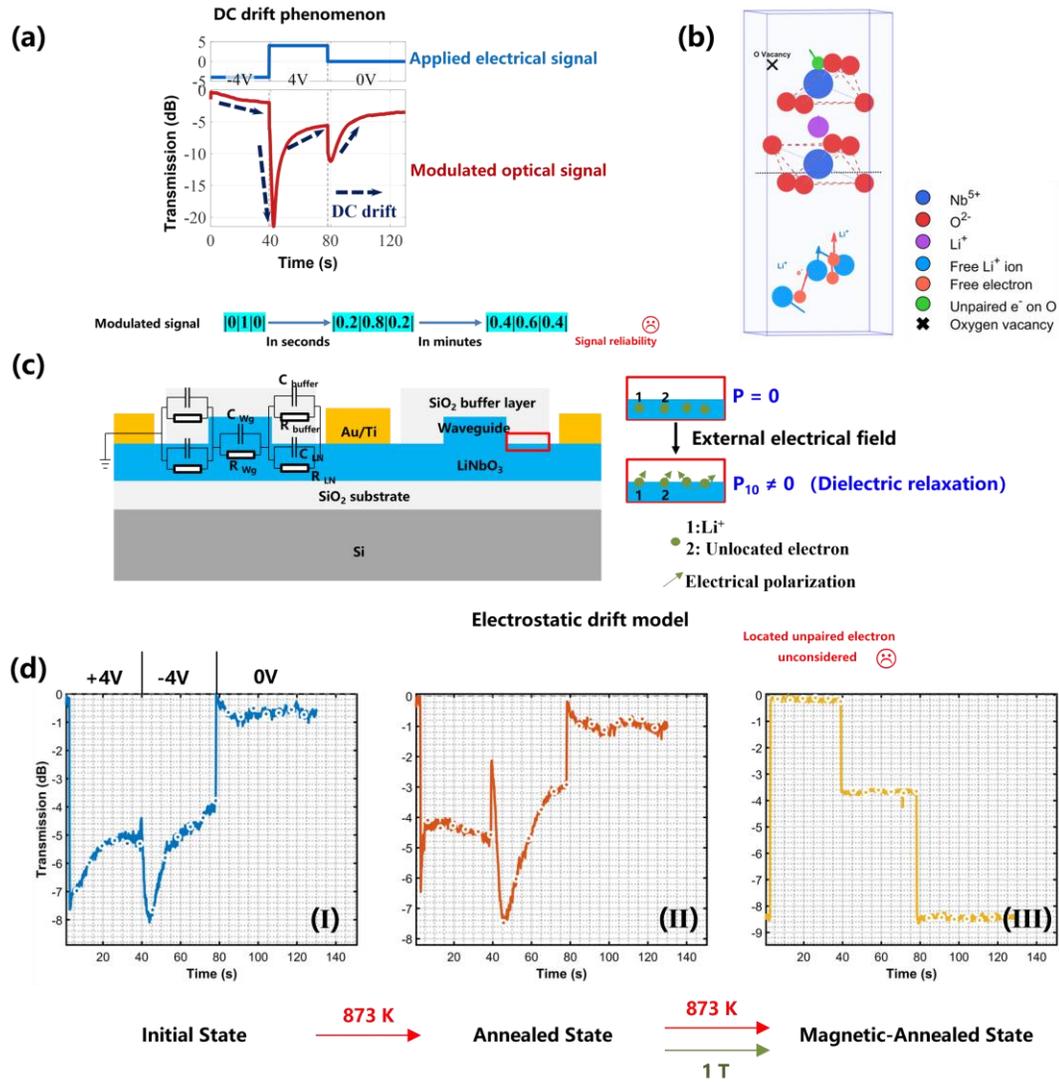

**Fig. 1. Suppression of DC drift in TFLN modulators. (a) Schematic of the DC drift phenomenon**: when a constant bias voltage is applied, the output optical signal intensity of the modulator fails to remain stable, continuously varying over time and eventually leading to complete signal distortion. **(b) Illustration of the possible origins of DC drift, covering two categories**: quasi-free electrons and Li$^+$ ions displaced from lattice sites; and unpaired electrons bound to atomic orbitals that produce a polarization response due to defects. **(c) Equivalent-capacitance model used to explain the DC drift phenomenon**: The LN layer and the SiO$_2$ cladding together form an RC loop, enabling the internal free charges to continuously alter their polarization response through circuit oscillation. **(e) DC Drift responses of the same modulator under different processing conditions:** The as-fabricated device shows significant drift, which is not stabilized by annealing alone. However, when an additional magnetic field is applied during annealing, the processed modulator exhibits strongly suppressed DC drift, and the optical signal becomes markedly stable after bias application. Note: All measurements were performed under ambient conditions (room temperature, atmospheric pressure, and without any external magnetic field).

In order to introduce the mEO effect, we need to first reinterpret the mechanism of the unresolved DC drift phenomenon found in LN/LT, although it has been widely recognized as a simple electrical relaxation in the past decades *(12-16)*. The DC drift phenomenon refers to the instability of the refractive-index change induced by an applied electric field in LN/LT materials; instead of remaining constant, the induced modulation continues to evolve over time (which can extend to several days), as illustrated in Fig. 1a. This behavior prevents active devices fabricated on such material platforms from maintaining stable phase control via a fixed voltage bias, leading to distortion of the optical signal during processing. It is widely recognized that DC drift originates from an anomalous polarization response within the material. This response is not intrinsic to the perfect crystal lattice but arises from charged particles that can produce additional electrical polarization. These particles can be broadly categorized into two types (Fig. 1b): The first category consists of mobile charges that have escaped the lattice confinement, primarily including displaced $Li^+$ ions and quasi-free electrons. These charges can be regarded as moving in a spatially averaged lattice potential; their states are not constrained by the band structure, and spin-related interactions are weak. The second category involves unpaired electrons associated with material defects. These electrons remain localized on atomic orbitals, preserving their polarization response while exhibiting well-defined band-structure characteristics. They participate in spin exchange/superexchange interactions and spin–orbit coupling, and can be excited via defect states, thereby influencing the ferroelectric response of the system *(17-19)*.

Previous studies on the mechanism of DC drift have predominantly focused on the first category, both theoretically and experimentally. A common feature of these works is that their theoretical modeling or experimental designs primarily consider the density distribution of charged carriers and the material conductivity. By incorporating these factors into numerical simulations via an equivalent-capacitance model (Fig. 1c), researchers found that removing the upper $SiO_2$ cladding of LN-based devices could interrupt the dielectric-relaxation loop and thereby reduce drift *(13)*. Although this leaves the optical waveguide exposed and compromises device stability, the approach proved effective. Furthermore, by employing different etching schemes and interface treatments to reduce the density of charged particles at interfaces, significant suppression of DC drift in LN has also been demonstrated *(14-16)*. However, to date, none of these strategies—individually or in combination—has succeeded in suppressing DC drift to an acceptable level; substantial operational-point drift still occurs within a short time after bias application.

Here, we present a straightforward experiment that demonstrates the prevailing understanding of the drift mechanism is incomplete. Specifically, we show that DC drift arises mainly from the second category: unpaired electrons associated with defects that remain bound within the atomic potential, rather than free carriers.

In the experiment, we performed multi-step processing on a typical MZI phase modulator on the LNOI platform and measured its DC drift after every step. The device was not processed according to previously accepted drift-suppression methods—neither the upper cladding was removed nor was additional ambient annealing performed (Its main design parameters will be listed at the end of the article). At the initial state, after applying a step bias (a switching sequence of 0 V → +4 V → −4 V → 0 V was used uniformly in our experiments), the optical signal intensity exhibited strong modulation but failed to stabilize, showing large variations over a short period (Fig. 1d(I)). We then subjected the same modulator to simple high-temperature annealing (873 K) and measured its drift behavior. The results showed that high-temperature annealing alone did not suppress the drift (Fig. 1d(II)), which is unsurprising given that LN has a ferroelectric Curie temperature as high as ~1410 K. However, when a magnetic field of 1 T was applied simultaneously during annealing, the outcome changed dramatically. Fig. 1d(III) displays the drift test result after the same modulator underwent further annealing under a 1 T magnetic field (note that the field was applied only during the treatment and was not maintained afterward). Clearly, after magnetic-field annealing, the DC drift of the device decreased substantially compared with its initial state or after annealing alone; almost no signal distortion was observed within 40 s after the voltage steps.

Evidently, the application of an external magnetic field does not interrupt the dielectric-relaxation loop, nor does it affect the population of delocalized $Li^+$ ions or quasi-free electrons. Therefore, this experiment directly demonstrates that the charged particles responsible for DC drift must possess a Hamiltonian that includes magnetic-coupling interactions, implying they remain bound within the lattice-forming atoms—i.e., they originate from the electric-field response of defect-related unpaired electrons. Moreover, because free carriers of the first category cannot exhibit non-volatile responses to an external magnetic field in the absence of a driving current, and given the substantial degree of drift suppression achieved, we conclude that DC drift in LN is predominantly caused by the second category.

It is worth noting that the observed reduction in drift after cladding removal can still be explained by the equivalent-capacitance model, because these unpaired electrons themselves possess a dipole moment, and their response to an electric field is analogous to that of free electrons; thus, interrupting the relaxation loop remains effective. Similarly, various interface treatments can concurrently alter interface defect states, thereby also mitigating the drift response associated with unpaired electrons to some extent.

## Mechanism of DC drift and the solution

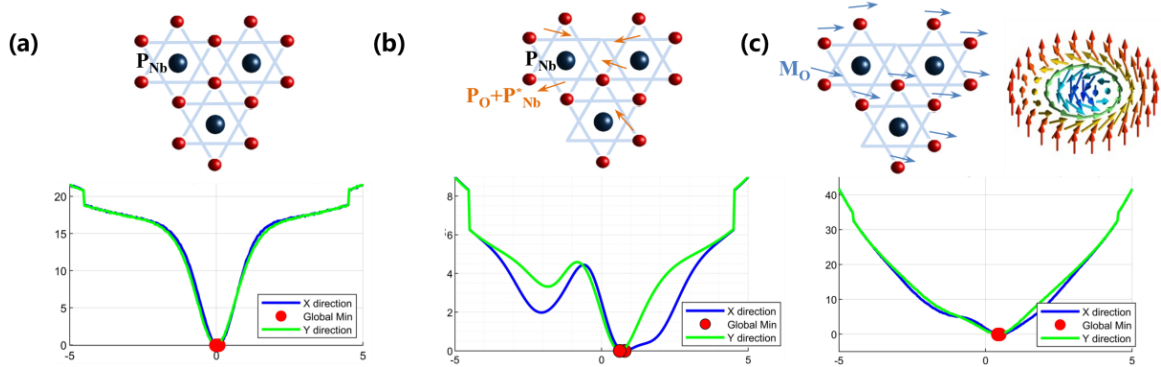

**Fig. 2. Origin and suppression mechanism of DC drift. (a) In an ideal LN lattice**: the primary polarizable electrons originate from the outer shells of Nb and Li ions. Under high crystallographic symmetry, these electrons exhibit a well-defined and prompt polarization response to an applied electric field. **(b) In a defective lattice**: the local symmetry breaking introduces additional interactions (e.g., ferromagnetic ordering) for the outer-shell electrons of Nb and Li. Furthermore, unpaired electrons associated with oxygen defects can provide an extra polarization contribution, either by modulating the behavior of intrinsic polarizable electrons or by responding directly to the field, with their binding reduced. These complex and varied interactions create a rugged energy landscape with multiple local metastable states for the electron polarization. **(c) LN lattice after skyrmion excitation**: Since both the Dzyaloshinskii–Moriya interaction (DMI) and the magnetic moments originate from the defect-related electrons, the formation of a skyrmion lattice necessarily imposes a collective magnetic order on these electrons. This order, protected by its topological nature, strongly couples the polarization response of individual electrons. Electrons that would otherwise be trapped in metastable states are now collectively "dragged" into a coherent polarization response by their neighbors, thereby drastically reducing the DC drift.

Having established that the DC drift in LN originates from the polar response of unpaired charges bound within the lattice, we now consider the physical mechanism underlying its relaxation behavior. Fundamentally, this phenomenon shares the same origin as the intrinsic EO effect of LN—both arise from the displacement of electrons that remain bound to atomic sites. The difference lies in the specific contributors: while the intrinsic EO coefficient primarily stems from the outer-shell electrons of Nb and Li, the drift response may receive additional contributions from, for instance, oxygen-related electrons. However, a key distinction exists in the response environment of these two "electro-optic" effects. Electrons occupying atomic orbitals within a lattice get stabilized with complex interactions—including exchange interaction and spin–orbit coupling—yet in a highly symmetric environment these interactions can be treated in a simplified manner. As an illustration, the free energy of an electron contributing to the intrinsic EO effect in LN can be expressed as a function of the order parameter x (representing the electron displacement):

$$F(x) = k_2 x^2 + k_4 x^4 + \cdots$$

Due to symmetry constraints, only even-order terms are present. This allows the polarizable electrons in LN to respond rapidly to an electric field and settle into a single, well-defined ground state (Fig. 2a). In contrast, the symmetry breaking induced by defects—which is often complex and random—places the field-responsive unpaired electrons in a much more complicated interaction environment that is difficult to capture accurately by first principles calculation. Nevertheless, a qualitative analysis remains feasible. In the presence of defects, symmetry breaking can be effectively described by introducing odd-order terms into the free-energy expansion:

$$F(x) = k_1 x + k_2' x^2 + k_3 x^3 + k_4' x^4 + \cdots$$

In such a scenario, the electron's response to an external electric field can exhibit multiple local energy minima (Fig. 2b). Consequently, under an applied field, an electron may become trapped in a random local metastable state rather than reaching the global minimum. Through thermal fluctuations or quantum tunneling, these electrons can stochastically hop among metastable configurations until eventually relaxing toward the global optimum, resulting in a slowly varying macroscopic polarization. This process cannot complete on short timescales but gradually decays as more electrons approach the global minimum, explaining why DC drift persists over long durations and its magnitude decays with time. Therefore, a viable route to suppress DC drift is to impose an additional, highly symmetric constraint that restores a simplified response landscape for these electrons. This is precisely what we achieve through the introduction of a room-temperature stable magnetic skyrmion state, as we previously reported *(20)*. The topologically protected skyrmion texture imposes strong collective order on the spin orientation and spatial arrangement of the unpaired electrons, forcing them to respond to external fields in a unified, coherent manner. This strong spin-spin interaction introduces an additional even-order potential into the free energy:

$$F(x) = k_1 x + k_2'' x^2 + k_3 x^3 + k_4'' x^4 + \cdots$$

Where $k_2''$ or $k_4'' \gg k_1$ and $k_3$. The emergence of this collective order dramatically reduces the influence of the local symmetry breaking, enabling the unpaired electrons to respond to an electric field in a manner analogous to the intrinsic polarization response of the perfect lattice (Fig. 2c). As a testable consequence of the proposed drift and suppression mechanisms, we derive the following corollary: if the response speed of unpaired electrons to an external field is significantly enhanced upon drift suppression, the LN material should exhibit an additional contribution to the electro-optic coefficient originating precisely from these now-ordered unpaired electrons. Remarkably, we indeed observe this phenomenon in devices after drift suppression.

Fig.3a(I) shows the voltage-dependent phase (VP) response of a modulator before drift suppression. Due to strong DC drift, the response is heavily distorted, preventing reliable extraction of the modulation efficiency. A separate measurement at 1 MHz yields a modulation efficiency of 2.04 V·cm (Fig. 3a(II)). After applying our drift-suppression treatment, the VP response becomes highly regular, with nearly identical voltage periods across cycles (Fig. 3b(I)), corresponding to a modulation efficiency of ~1.4 V·cm. This improvement is confirmed by the 1 MHz measurement (Fig. 3b(II)). Since no structural phase transition occurs in LN, the additional electro-optic contribution cannot arise from the intrinsic ferroelectric response.

Therefore, we can finally conclude that the true origin of DC drift lies in defect-related unpaired electrons, and its long-duration relaxation stems from symmetry breaking induced by defects. The suppression of DC drift via magnetic-field annealing essentially relies on the topological protection of skyrmions to impose an additional collective constraint on the unpaired electrons, thereby drastically mitigating the influence of symmetry-broken metastable states.

## Magnetic Control of the mEO Effect

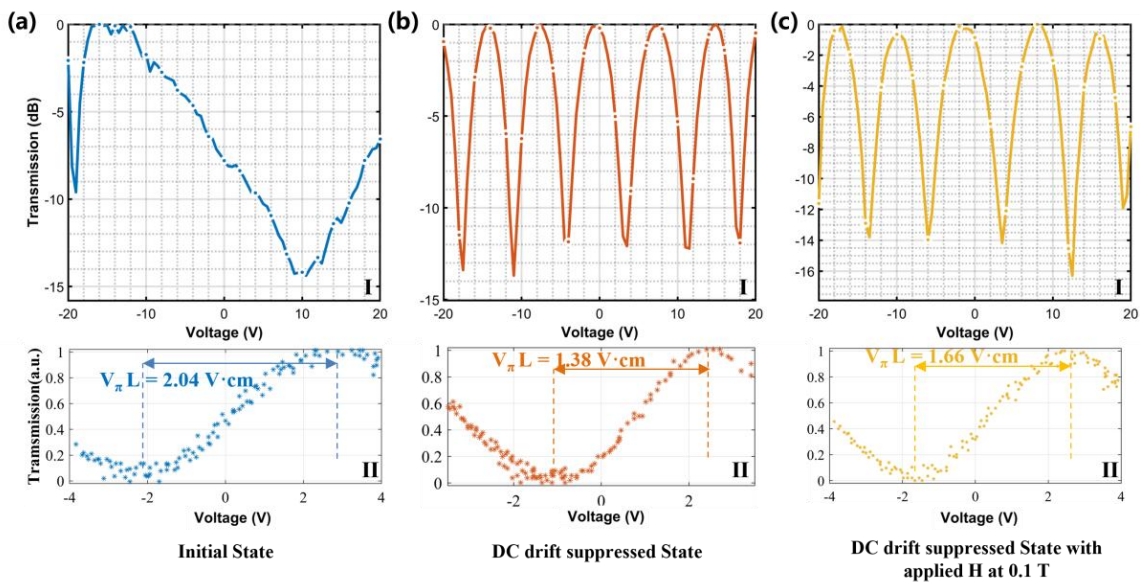

**Fig. 3. Observation and magnetic-field response of the magnetically coupled electro-optic effect. (a) Initial state of the test MZI modulator: (I)** Low-frequency voltage-phase (VP) scan showing a severely distorted and irregular response, indicating ineffective modulation due to strong DC drift. **(II)** Modulation performance measured at 1 MHz, yielding an efficiency of 2.04 V·cm. **(b) MZI modulator (a) after DC drift suppression: (I)** Low-frequency VP scan revealing a regular, uniform response with successful modulation. The extracted half-wave voltage is significantly higher than the simulated value, indicating the presence of an additional electro-optic contribution. **(II)** Modulation efficiency at 1 MHz, improved to 1.38 V·cm. **(c) MZI modulator (b) with an applied magnetic field of 0.1 T along the x-axis of the LN crystal. (I)** Low-frequency VP scan maintains a regular line shape suitable for modulation, though with slightly reduced uniformity compared to **(b, I)**. The half-wave voltage remains substantially higher than the simulation. **(II)** Modulation efficiency at 1 MHz, modulated to 1.66 V·cm upon magnetic field application, directly demonstrating the tunability of the mEO effect.

Given that this pronounced electro-optic effect is modulated by magnetic skyrmions, it is highly plausible that it also exhibits magnetic-field responsiveness—i.e., it enables the control of a material's polarization response to an external electric field via a magnetic field (the magnetically coupled electro-optic effect). We therefore further investigated the magnetic-coupling capability of this electro-optic contribution. Following the same measurement protocol as before, we applied an extra magnetic field of 0.1 T to a modulator whose drift had been suppressed and which exhibited the magnetically induced electro-optic effect. A clear magnetoelectric response was observed. Under the applied field, the VP curve broadened significantly, and the half-wave voltage increased substantially—indicating strong suppression of the electro-optic effect. The same magnetic-field response was confirmed by modulation-efficiency measurements at 1 MHz. To describe the magnetoelectric coupling formally, we consider the free-energy expansion *(11)*:

$$F(E,H) = -\frac{1}{2}\epsilon_0 x^e_{ij} E_i E_j - \frac{1}{2}\mu_0 x^m_{ij} H_i H_j - \alpha_{ij} E_i H_j - \frac{1}{2}\beta_{ijk} E_i H_j H_k - \frac{1}{2}\gamma_{ijk} E_i E_j H_k$$

where the first two terms represent the electric and magnetic polarization energies, the third term corresponds to the linear magnetoelectric coupling, and the last two terms describe higher-order magnetoelectric contributions. The magnetic tuning of the electro-optic strength belongs to the final term, in which the electric susceptibility $x^e_{ij}$ acquires a field-dependent correction proportional to $\gamma_{ijk} H_k$. The modulation efficiency of 1.66 V·cm measured after the applied magnetic field means that under the current drift processing process, LN can obtain an additional mEO coefficient of nearly $8*10^{-3} pm/V/Oe$, and the electro-optical coefficient of 8 pm/V can be adjusted under a magnetic field of 0.1T, which is equivalent to 27% of the EO coefficient of LN itself. This observation is remarkable for two reasons. First, previously reported magnetoelectric couplings rarely persist at room temperature, and even those that do typically exhibit very weak signals. Second, most studies on strong magnetoelectric effects focus on the linear regime—either electric-field control of magnetization or magnetic-field control of electric polarization *(1-3)*—because higher-order magnetoelectric couplings are generally much weaker and subject to far more stringent symmetry constraints. Given that linear magnetoelectric materials themselves are scarce and challenging to realize, the search for appreciable higher-order magnetoelectric effects has been considered extremely difficult and of limited practical relevance. Our study, however, demonstrates for the first time that a higher-order magnetoelectric coupling can not only be remarkably strong but also operate robustly at room temperature.

## MEO effects in LT

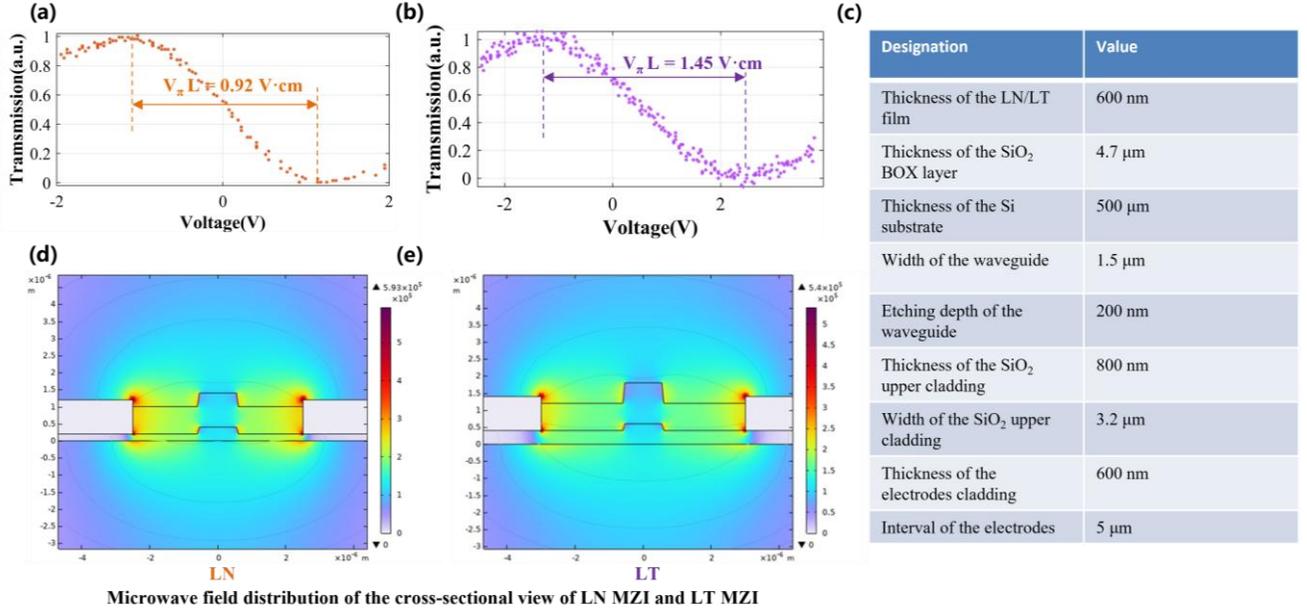

Microwave field distribution of the cross-sectional view of LN MZI and LT MZI

**Fig. 4. Enhanced electro-optic response in LN and LT after skyrmion stabilization.**
**(a, b)** Measured modulation efficiency at 1 MHz for the **(a)** LN and **(b)** LT modulators, demonstrating an enhanced EO coefficient following optimization.
**(c)** Key design parameters of the implemented MZI modulators.
**(d, e)** Simulated cross-sectional distribution of the modulating electric field within the waveguide active region for the **(d)** LN and **(e)** LT devices, confirming efficient field overlap with the optical mode.

In this work, we have proposed a new interpretation of the physical origin of DC drift and validated it through drift-suppression experiments in LN and the demonstration of a magnetically coupled electro-optic effect. A direct and critical extension of this finding is to examine whether LT—which also exhibits DC drift—can manifest the same phenomena, i.e., an additional magnetically coupled electro-optic response. The answer is unequivocally affirmative. Fig 4(a) presents the experimentally achieved electro-optic coefficients for both LN and LT under optimized conditions. LN exhibits coefficients of 66 pm/V, while LT shows a pronounced enhancement reaching approximately 45 pm/V. This difference aligns with the previously reported observation that the intrinsic DC drift in LT is generally weaker than that in LN, further corroborating our proposed mechanism for the drift origin and the source of the additional electro-optic contribution. To substantiate the authenticity of the magnetically induced electro-optic effect in both material systems, we provide the key design parameters of the LN and LT modulators used in our experiments, along with simulated profiles of the optical mode and the applied electric field. These confirm that the observed enhancements are indeed attributable to the modified material response rather than variations in device geometry or field overlap.

## Conclusion

In conclusion, we have proposed and experimentally validated a new interpretation of the DC drift mechanism in LN and LT, revealing its fundamental nature as a room-temperature magnetoelectric coupling phenomenon arising from defect-bound unpaired electrons. Furthermore, by deliberately suppressing the drift, we have unlocked and stabilized a previously unreported magnetically coupled electro-optic effect at room temperature, and quantitatively characterized its magnetoelectric response. Our work resolves a decades-long debate concerning the origin of DC drift in these technologically crucial materials and provides a practical solution for its suppression. These findings pave the way for the reliable deployment of high-performance ferroelectric optical materials like LN and LT in next-generation photonic integrated circuits, optical interconnects, and high-speed communication systems. More broadly, this study opens a new avenue for exploring magnetoelectric phenomena and their associated material platforms, demonstrating the significant potential of tailored spin-electronic states for practical device engineering.